\documentclass[runningheads]{llncs}
\pdfoutput=1 

\usepackage{appendix}
\usepackage{booktabs}
\usepackage[T1]{fontenc}
\usepackage{graphicx}
\usepackage{hyperref}
\usepackage{multirow}
\usepackage{tabularx}
\RequirePackage{pgfplots}
\pgfplotsset{compat=1.18}
\RequirePackage{tikz}

\usepackage{color}

\newcommand{\ra}[1]{\renewcommand{\arraystretch}{#1}}

\begin{document}

\title{Domain Influence in MRI Medical Image Segmentation: spatial versus k-space inputs}

\titlerunning{Domain Influence in MRI Medical Image Segmentation}

\author{Erik Gösche\inst{1}\orcidID{0009-0009-7547-3746} \and
Reza Eghbali\inst{2,3}\orcidID{0000-0003-4856-3059} \and
Florian Knoll\inst{1}\orcidID{0000-0001-5357-8656} \and
Andreas M Rauschecker\inst{3}\orcidID{0000-0003-0633-9876}
}

\authorrunning{E. Gösche et al.}

\institute{Department of Artificial Intelligence in Biomedical Engineering, University of Erlangen–Nuremberg, Erlangen, Germany\\ 
\email{erik.goesche@fau.de, florian.knoll@fau.de} \and
Berkeley Institute for Data Science (BIDS), University of California, Berkeley,
Berkeley, CA 94720, USA\\
\email{eghbali@berkeley.edu} \and
Department of Radiology and Biomedical Imaging, University of California, San Francisco,
San Francisco, CA 94158, USA\\
\email{andreas.rauschecker@ucsf.edu}
}

\hypersetup{pdfauthor={Erik Gösche, Reza Eghbali, Andreas Rauschecker, Florian Knoll},pdftitle={Domain Influence in MRI Medical Image
Segmentation: spatial versus k-space inputs}}

\maketitle

\begin{abstract}
Transformer-based networks applied to image patches have achieved cutting-edge performance in many vision tasks. However, lacking the built-in bias of convolutional neural networks (CNN) for local image statistics, they require large datasets and modifications to capture relationships between patches, especially in segmentation tasks. Images in the frequency domain might be more suitable for the attention mechanism, as local features are represented globally. By transforming images into the frequency domain, local features are represented globally. Due to MRI data acquisition properties, these images are particularly suitable. This work investigates how the image domain (spatial or k-space) affects segmentation results of deep learning (DL) models, focusing on attention-based networks and other non-convolutional models based on MLPs. We also examine the necessity of additional positional encoding for Transformer-based networks when input images are in the frequency domain. For evaluation, we pose a skull stripping task and a brain tissue segmentation task. The attention-based models used are PerceiverIO and a vanilla Transformer encoder. To compare with non-attention-based models, an MLP and ResMLP are also trained and tested. Results are compared with the Swin-Unet, the state-of-the-art medical image segmentation model. Experimental results indicate that using k-space for the input domain can significantly improve segmentation results. Also, additional positional encoding does not seem beneficial for attention-based networks if the input is in the frequency domain. Although none of the models matched the Swin-Unet's performance, the less complex models showed promising improvements with a different domain choice.

\keywords{Medical Image Segmentation  \and Frequency Domain Analysis \and Attention-based Networks.}
\end{abstract}

\section{Introduction}
The success of the Transformer model~\cite{vaswani_attention_2017} and later the Vision Transformer~\cite{dosovitskiy_image_2020} has led to various attention-based models achieving state-of-the-art performance in medical vision tasks~\cite{lin_seg4reg_2022,hatamizadeh_unetr_2022,gutsche_automated_2023}. The attention mechanism enables an excellent way to capture long-range dependencies within the input, which has been shown to be an advantage in vision tasks~\cite{zhang_graph_2022}. 
However, Transformer-based networks struggle with large-scale inputs like medical images due to their quadratic complexity with respect to input size~\cite{vaswani_attention_2017}. To address this issue, input images are typically subdivided into patches. Training of Transformer-based models requires large amounts of data to capture relationships among these image patches. The reason for this comes from the lack of inherent biases in Transformer models, such as local receptive fields or shared weights, which facilitate feature learning in CNNs. Particularly for pixel data, the explicit correlation of adjacent pixels, often strongly correlated, and translation in-variance are desirable characteristics~\cite{wu_cvt_2021}. An additional consequence of the absent receptive field is that every input value contributes to attention computation, resulting in interrelated values. While this aids the extraction of global features, it makes it difficult to capture local features~\cite{chen_transunet_2021}.

Enhancing attention-based models can be achieved by using images in the frequency domain. Converting images to the frequency domain allows local features to be represented globally, leveraging the attention mechanism's ability to capture long-range dependencies. This approach is particularly relevant for medical imaging tasks, such as magnetic resonance imaging (MRI), where the properties of the data acquisition process make frequency domain representations particularly suitable. The convolution theorem, stating that convolution in image space is equivalent to element-wise product in Fourier space, inspired this method. This means that the convolutional layer in the image space can be replaced by a simple linear projection layer where each neuron applies a weight to one of the Fourier coefficients in the frequency space. The attention layer is a special case of a fully connected layer. However, the existence of non-linearities prevents us from further extending this line of thought. Furthermore, for natural language processing (NLP) tasks, it has been observed that in the frequency domain, additional position coding for input data, which is typically required for attention operations, may be unnecessary~\cite{lee-thorp_fnet_2022}. This research investigates how the choice of image domain (spatial or k-space) affects segmentation results of DL models, focusing on simple models, including attention-based networks and other non-convolutional networks based on multilayer perceptrons (MLPs). Additionally, it examines the necessity of additional positional encoding for Transformer-based networks when input images are in the frequency domain. 

To address these research questions, two segmentation tasks are posed: skull stripping and brain tissue segmentation. The attention-based models evaluated in this study include the PerceiverIO~\cite{jaegle_perceiver_2022} and a vanilla Transformer encoder. To provide a comprehensive comparison, non-attention-based models such as an MLP and the ResMLP~\cite{touvron_resmlp_2021} are also trained and tested. The performance of these models is compared with that of the Swin-Unet~\cite{karlinsky_swin-unet_2023}, a state-of-the-art medical image segmentation model.

\subsection{Related work}
Several authors have proposed DL models for image classification and reconstruction tasks that utilize the representation of images in the frequency domain~\cite{jiang2021focal,stuchi2020frequency,zhou2022deep,rao2021global}. Recently, Wang et al. proposed a model for lesion segmentation in brain MRIs that uses masked image modeling in the frequency domain as a self-supervised pre-training stage~\cite{wang2024fremim}. These models generally include sub-blocks that consist of a 2D Fast Fourier Transform (FFT) layer, a specific processing operation in the frequency domain, and a 2D inverse FFT to transform the data back to the spatial domain. Since our goal is to compare the performance of Transformer and MLP based architectures across different domains, we do not include any FFT or inverse FFT layer inside our models and keep the model architecture consistent across different domains. 

Another related line of research is machine learning-based undersampled MRI reconstruction (see~\cite{hammernik2021systematic,singh2023emerging} and references therein) and simultaneous reconstruction and segmentation~\cite{tolpadi2023k2s,huang2019brain}. This is inherently a cross-domain problem where the network has to predict spatial space values from k-space samples. These methods most often use convolutional layers, variational networks, and recently transformers~\cite{zhao2022k}. However, segmentation is done on the reconstructed image in the spatial domain. 

\section{Method}
\subsection{Model Architectures}
In this work, we employ the PerceiverIO and a vanilla Transformer encoder as exemplars of attention-based models. Additionally, we incorporate an MLP and the ResMLP for comparative analysis. Convolution-based models are not selected due to their incompatibility with the frequency domain, where each point represents information across the entire spatial domain, making these models misleading. This selection involves relatively simple models in terms of their complexity. The following experiments can therefore be used to determine how the lack of model capacity can be compensated by selecting suitable domains. To enable further evaluation of the results, widely used models such as nnU-Net~\cite{isensee_nnu-net_2021} and Swin-Unet, the latter of which is also recognized as state of the art, are also included.
The MLP is the simplest model among the four, serving as a proof-of-concept. It consists of linear input and output embeddings, N hidden fully connected layers, each followed by a tanh activation layer. Input reshaping reduces complexity, and both the dimensionality (M) of the latent space and the number of hidden layers (N) are hyperparameters. Similarly, the Transformer encoder employs linear embeddings for input and output, incorporating N encoder components from the Transformer model. The architecture resembles BERT~\cite{devlin_bert_2019}, with the number of encoder blocks and latent space dimensionality being hyperparameters. For the implementation of the Transformer encoder, the corresponding PyTorch class was used. Fourier position encoding may be concatenated with the input. The PerceiverIO, implemented using Krasser and Stumpf's Python module~\cite{krasser_pytorch_2023}, utilizes cross-attention to map input to a smaller latent space, reducing attention complexity from quadratic to linear. The ResMLP model, following documentation, resembles the vision transformer but lacks attention layers. Linear layers replace attention layers, and traditional normalizations are omitted in favor of affine transformations. For this work, the ResMLP was implemented so that the input is not divided into patches and embedded using a linear projection. Instead, the sagittal slices of the MRI brain data serve as channels and the remaining dimensions are flattened.

\section{Experiments and Results}
\subsection{Datasets}
We present two segmentation tasks, each accompanied by its own dataset. For the initial task of skull stripping, we employ the UPENN-GBM dataset~\cite{bakas_university_2022}. As a follow-up task, focusing on brain tissue segmentation, we aim to highlight variations in complexity levels across segmentation tasks, utilizing the OASIS-1 dataset~\cite{marcus_open_2007}. Both datasets are freely available to ensure the reproducibility of this work. Follow-up scans in both datasets are excluded from this work. For the brain tissue segmentation task, only the OASIS FreeSurfer output (brain mask as input and tissue segmentation for labels) is used. The number of tissue segmentation classes have been simplified to six classes (cortical gray matter, white matter, CSF, deep gray matter, brain stem and cerebellum). The exact mapping can be seen in the published source code. This way, 611 subject from the UPENN-GBM dataset and 407 subjects from the OASIS-1 dataset are used. All samples are converted to NIfTI format and sampled to an isotropic voxel size of 3 mm to reduce complexity. Furthermore, the samples are cropped to a size of $64\times{}64\times{}64$ and z-normalized. We apply the subsequent augmentation transformations: random affine transformation, random contrast adjustment, random Gaussian noise addition, random MRI motion artifact introduction, and random MRI bias field artifact inclusion. For both segmentation tasks, we partitioned the dataset into three subsets: training (80\%), validation (10\%), and testing (10\%).

\subsection{Implementation Details}
This study uses Python (version 3.10.12), PyTorch (version 2.0.1), TorchIO (version 0.18.92), Ray (version 2.5.1), and PyTorch Lightning (version 2.0.5) for all implementations and analyses~\footnote{The source code is publicly available at \url{https://www.github.com/rauschecker-sugrue-labs/kspace-segmentation}.} To propagate the slices of each sample through the network in the frequency domain, we employ the 2D real FFT implementation in PyTorch. This method operates under the assumption that the outcome of the FFT is Hermitian symmetric, which holds true in our case since we are using already reconstructed MRI data. By exploiting this symmetry, the input is halved. As the real part and imaginary part of the resulting complex numbers are saved separately in two vectors, the size of the input remains the same even after the Fourier transformation.
The training routine for all non-baseline models is designed to allow independent specification of the target domain for inputs and labels at startup. This flexibility enables the definition of various combinations of input and label domains. When labels are in the spatial domain, the task becomes a classification problem and the model is trained using a cross-entropy loss; when labels are transformed into the frequency domain by using the 2D FFT, it becomes a regression problem and the model is trained using mean squared error loss. We use three domain configurations. In the spatial domain, both input and output are in the spatial domain. In the k-space domain, the input is transformed into the frequency domain using FFT, and the output is predicted in the frequency domain. In the k-space-to-spatial domain, the input is in the frequency domain, and the output is predicted in the spatial domain.

\subsection{Segmentation Performance in different Domains}
This section presents quantitative segmentation results from all experiments. Public source code records hyperparameter configurations used for obtaining these results. The results of the skull stripping task among the models are displayed in Table~\ref{tab:ss_domains}. Notably, in the spatial domain, Dice scores are similar across models, with MLP slightly underperforming in recall and specificity. In k-space, MLP performs worse than in the spatial domain, while other models perform similarly. MLP remains weakest in the k-space domain.

\begin{table*}
\centering
\caption{Dice similarity coefficient (DSC), sensitivity (sens) and specificity (spec) for different models on skull-stripping in spatial domain, k-space and k-space-to-spatial domain.}
\ra{1.3}
\begin{tabular}{@{}r @{\hspace{5.5pt}} r @{\hspace{5.5pt}} r @{\hspace{5.5pt}} r c r @{\hspace{5.5pt}} r @{\hspace{5.5pt}} r c r @{\hspace{5.5pt}} r @{\hspace{5.5pt}} r @{\hspace{5.5pt}} r@{}}
\toprule
& \multicolumn{3}{c}{Spatial} & \phantom{abc}& \multicolumn{3}{c}{K-Space} &
\phantom{abc} & \multicolumn{3}{c}{K-Space to Spatial}\\
\cmidrule{2-4} \cmidrule{6-8} \cmidrule{10-12}
& DSC & Sens & Spec && DSC & Sens & Spec && DSC & Sens & Spec\\ \midrule
MLP & 0.964 & 0.960 & 0.992 && 0.898 & 0.888 & 0.976 && 0.966 & 0.968 & 0.990\\
ResMLP & \textbf{0.978} & 0.973 & 0.995 && \textbf{0.978} & 0.980 & 0.993 && \textbf{0.976} & 0.977 & 0.993\\
PerceiverIO & 0.930 & 0.923 & 0.984 && 0.927 & 0.919 & 0.984 && 0.929 & 0.919 & 0.985\\
Trans. Encoder & 0.971 & 0.971 & 0.992 && 0.971 & 0.969 & 0.993 && 0.972 & 0.973 & 0.993\\
\midrule
nnU-Net & 0.986 & 0.987 & 0.996 && & & && & & \\
Swin-UNet & 0.994 & 0.995 & 0.974 && & & && & & \\
\bottomrule
\end{tabular}
\label{tab:ss_domains}
\end{table*}

For brain tissue segmentation, Fig.~\ref{fig:tissue_domains} displays results for spatial, k-space, and k-space-to-spatial domains, respectively. Detailed evaluations can be found in the supplementary material. There are notable differences in model performance in contrast to to skull stripping. Starting in spatial domain, ResMLP outperforms others non-baseline models significantly with a Dice score of 0.876, followed by the Transformer encoder and the MLP. This means, the Transformer encoder surpasses MLP by approximately 12\% in spatial domain. PerceiverIO performs poorly with a Dice score of only 0.418. In k-space, ResMLP's performance declines to a Dice score of 0.815, while other models improve. The ResMLP is no longer able to detect the fine structures of the cortical gray matter and therefore classifies these areas as too large (see Fig.~\ref{fig:tissue}). The Transformer encoder notably improves by about 13\% to a Dice score of 0.790. PerceiverIO's performance for this class slightly improves with the domain change to a score of 0.461. The MLP achieves an improved Dice score of 0.690. In the k-space-to-spatial domain, ResMLP excels, achieving a Dice score of 0.883. All models benefit from this domain regarding cortical gray matter segmentation except PerceiverIO. MLP's performance slightly worsens compared to k-space but improves compared to the spatial domain. The Transformer encoder achieves its best performance in the k-space-to-spatial domain with a Dice score of 0.861. As it can be seen in the displayed segmentation mask, the Transformer encoder was able to delineate the individual classes more sharply.
\begin{figure}
    \centering
    \begin{tikzpicture}
    \begin{axis}[
    xbar,
    bar width=0.2,
    ytick={1,2,3,4,5,6},
    yticklabels={Swin-Unet, nnU-Net, ResMLP, MLP, Trans. Encoder, PerceiverIO},
    yticklabel style={font=\small}, 
    xlabel={Dice score},
    legend style={
        at={(0.5, 1.1)}, 
        anchor=north, 
        legend columns=-1,
        legend cell align={left}, 
        column sep=5pt, 
    },
    width=11cm, 
    height=8cm, 
    xmajorgrids=true, 
    grid style={dashed,gray!30}, 
    nodes near coords style={font=\scriptsize},
    xmin=0.4, 
    xmax=1.02, 
    ]
    \addplot+[error bars/.cd,
    x dir=both,x explicit]
    coordinates {
        (0.912,1) +- (0.01, 0.0)
        (0.956,2) +- (0.01, 0.0)
        (0.876,3) +- (0.0184, 0.0)
        (0.625,4) +- (0.0331, 0.0)
        (0.698,5) +- (0.0308, 0.0)
        (0.418,6) +- (0.0775, 0.0)};
    \addplot+[error bars/.cd,
    x dir=both,x explicit]
    coordinates {
        (0.815,3) +- (0.0212, 0.0)
        (0.690,4) +- (0.0527, 0.0)
        (0.790,5) +- (0.0248, 0.0)
        (0.461,6) +- (0.1125, 0.0)};
    \addplot+[error bars/.cd,
    x dir=both,x explicit]
    coordinates {
        (0.883,3) +- (0.0211, 0.0)
        (0.676,4) +- (0.0292, 0.0)
        (0.861,5) +- (0.0221, 0.0)
        (0.419,6) +- (0.0810, 0.0)};

    \legend{Spatial, K-Space, K-Space $\rightarrow$ Spatial}
    \end{axis}
    \end{tikzpicture}
    \caption{Mean and standard deviations of the Dice score among all models on brain tissue segmentation in spatial, k-space, and k-space-to-spatial domain. The mean values are represented by the bars and the standard deviations are indicated by the error bars.}
    \label{fig:tissue_domains}
\end{figure}
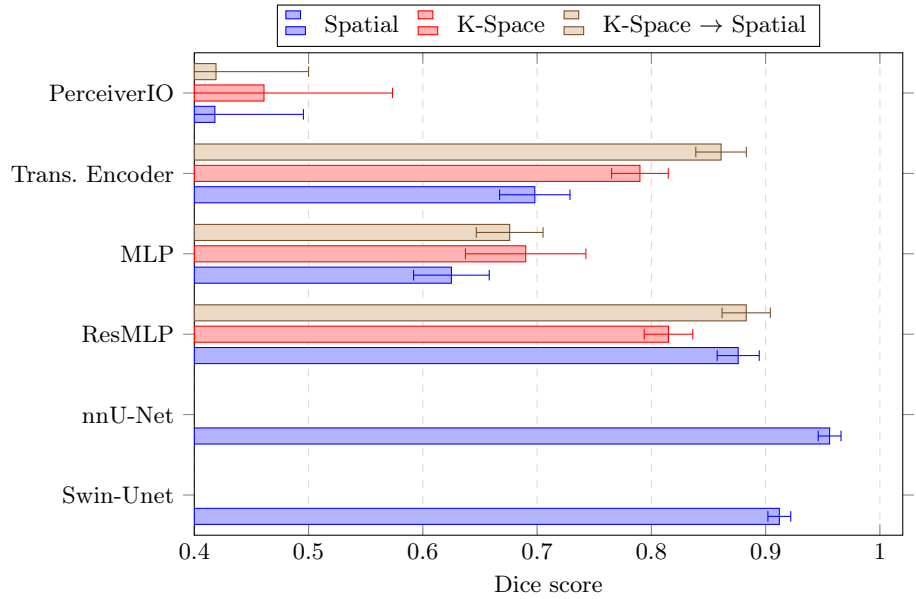
\begin{figure}[ht]
\centering
\includegraphics[width=\textwidth]{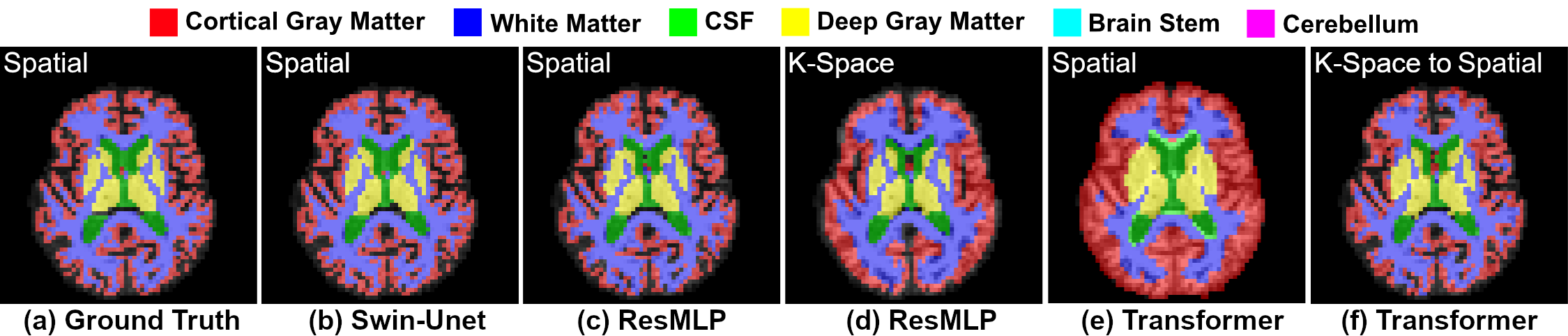}
\caption{Segmentation masks of the different models for brain tissue segmentation among varies domains.}
\label{fig:tissue}
\end{figure}

The following table~\ref{tab:pos_enc} presents the results regarding the necessity of additional position coding for attention-based networks when working with k-space input. The table outlines the comparison between PerceiverIO and Transformer encoding, with and without additional positional encoding, across three distinct domains for the skull stripping task and the brain tissue segmentation task. Notably, no substantial differences are discernible within each model, across all three metrics.

\begin{table}
  \centering
  \caption{Dice scores for attention-based models with and without positional encoding (PE) on skull stripping and brain tissue segmentation in k-space. Average Dice score is reported for brain tissue segmentation task.}
  \begin{tabular}{l>{\centering\arraybackslash}p{2cm} >{\centering\arraybackslash}p{2cm} >{\centering\arraybackslash}p{2cm} >{\centering\arraybackslash}p{2cm}}
    \toprule
    \multirow{2}{*}{Architecture} &
      \multicolumn{2}{c}{Skull Stripping} &
      \multicolumn{2}{c}{Brain Tissue Segmentation} \\
      & PE & no PE & PE & no PE \\
      \midrule
    PerceiverIO & 0.927 & 0.930 & 0.461 & 0.463 \\
    Trans. Encoder & 0.971 & 0.970 & 0.790 & 0.790 \\
    \bottomrule
  \end{tabular}
  \label{tab:pos_enc}
\end{table}

\subsection{Complexity comparison with State-of-the-Art Models}
The training and testing of the nnU-Net and Swin-Unet in spatial domain shows that both models outperform the other models in any domain constellation. However, in skull stripping the differences are relatively small. The reason for this is most likely because this task is not complex enough to see differences between domains. Looking at the results of the brain tissue segmentation in Fig.~\ref{fig:tissue_domains}, the differences become more obvious. In this analysis, it becomes clear that nnU-Net and the Swin-Unet exhibits the most effective performance in brain tissue segmentation. However, both the Transformer encoder in the k-space-to-spatial domain and ResMLP offer competitive segmentation outcomes. 
This becomes particularly clear when comparing the complexity of these models. Table~\ref{tab:flops} shows the floating point operations (FLOPs) required for the forward and backward pass of the models used. The number of model parameters is also documented. The results clearly show that although the Swin-Unet outperforms the other non-baseline models, it is also much more complex. Comparing the Transformer encoder with the Swin-Unet, it can be seen that more than twice as many FLOPs are required for the forward and backward pass. In addition, the ResMLP requires less then 600$\times{}10^9$ FLOPs for both passes. Taking this into consideration, the performance of the Transformer encoder and ResMLP is impressive.
In summary, the choice of domain significantly impacts the results for brain tissue segmentation. The correct domain can compensate for a lack of model capacity and thus good segmentation results can be achieved with simple models.

\begin{table}[htbp]
\centering
\caption{Comparison of FLOPs and parameters in different architectures (FFT/iFFT not included)}
\begin{tabular}{@{}l>{\centering\arraybackslash}p{2cm} >{\centering\arraybackslash}p{2cm} >{\centering\arraybackslash}p{2cm}}
\toprule
Architecture    & FLOPs (Forward) & FLOPs (Backward) & Parameters (M) \\ \midrule
MLP             & 128.85G          & 231.93G          & 83.91         \\
ResMLP          & 208.57G          & 391.38G          & 134.43        \\
Trans. Encoder  & 361.38G          & 696.99G          & 234.97        \\
PerceiverIO     & 943.30G          & 1,886.60G        & 47.54         \\
Swin-Unet       & 746.88G          & 1,491.20G        & 234.97        \\ \bottomrule
\end{tabular}
\label{tab:flops}
\end{table}

\section{Conclusion and Limitations}
This study examined DL model performance, especially attention-based and non-convolutional types, for brain segmentation tasks across domains, emphasizing both spatial and frequency domains. The study demonstrated how brain segmentation outcomes vary when input and label data are independently presented in either the spatial or frequency domain during supervised learning. Four models were implemented: PerceiverIO, a Transformer encoder, an MLP, and ResMLP, focusing on three domain configurations: spatial-to-spatial, k-space-to-k-space, and k-space-to-spatial. Skull stripping and brain tissue segmentation tasks were selected. Results indicated that domain configuration significantly impacts segmentation performance for sufficient complex tasks. For example, using the Transformer encoder, brain tissue segmentation performance improved by over 23\% when data was transformed into the frequency domain, measured by the Dice score. This supports the idea that Fourier-transformed input data is better suited for attention-based networks like the Transformer encoder. However, this was not observed for the skull stripping task, likely due to its simplicity. The Transformer encoder and ResMLP performed best in the k-space-to-spatial domain configuration, possibly due to easier prediction of segmentation masks in the spatial domain, which exhibit an imbalance in frequency components. Additionally, additional positional encoding is unnecessary when input data is in the frequency domain, extending findings from NLP tasks to computer vision. Results were compared with Swin-Unet, the baseline model. Despite Swin-Unet outperforming the implemented models, ResMLP showed competitive performance in the k-space-to-spatial domain, considering its relative simplicity.

Some aspects of this work limit the scope and applicability of the results obtained. One aspect is that only the binary cross-entropy loss, cross-entropy loss and mean squared error loss were considered as loss functions in this work. Other loss functions such as Dice Loss or Focal Loss would also be worth considering. The OASIS dataset is freely accessible and offers isotropic resolution, which simplifies our pre-processing pipeline. However, it should also be noted that this dataset contains presumably healthy subjects, which makes the segmentation task easier. Also, the approach used in this work cannot be directly translated into the clinical setting. By using the real 2D FFT it was assumed that the input is Hermitian symmetric. However, this assumption cannot be made for raw MRI k-space data.

\begin{credits}
\subsubsection{\ackname} We gratefully acknowledge the scientific support and HPC resources provided by the Erlangen National High Performance Computing Center (NHR @FAU) of the Friedrich-Alexander-Universität Erlangen-Nürnberg (FAU). The hardware is funded by the German Research Foundation (DFG).

\subsubsection{\discintname}
We have no competing interests.
\end{credits}

\bibliographystyle{splncs04}
\bibliography{literature}

\begin{thebibliography}{10}
\providecommand{\url}[1]{\texttt{#1}}
\providecommand{\urlprefix}{URL }
\providecommand{\doi}[1]{https://doi.org/#1}

\bibitem{bakas_university_2022}
Bakas, S., Sako, C., Akbari, H., Bilello, M., Sotiras, A., Shukla, G., Rudie, J.D., Santamaría, N.F., Kazerooni, A.F., Pati, S., Rathore, S., Mamourian, E., Ha, S.M., Parker, W., Doshi, J., Baid, U., Bergman, M., Binder, Z.A., Verma, R., Lustig, R.A., Desai, A.S., Bagley, S.J., Mourelatos, Z., Morrissette, J., Watt, C.D., Brem, S., Wolf, R.L., Melhem, E.R., Nasrallah, M.P., Mohan, S., O’Rourke, D.M., Davatzikos, C.: The {University} of {Pennsylvania} glioblastoma ({UPenn}-{GBM}) cohort: advanced {MRI}, clinical, genomics, \& radiomics. Scientific Data  \textbf{9}(1), ~453 (Jul 2022). \doi{10.1038/s41597-022-01560-7}, number: 1 Publisher: Nature Publishing Group

\bibitem{karlinsky_swin-unet_2023}
Cao, H., Wang, Y., Chen, J., Jiang, D., Zhang, X., Tian, Q., Wang, M.: Swin-{Unet}: {Unet}-{Like} {Pure} {Transformer} for {Medical} {Image} {Segmentation}. In: Karlinsky, L., Michaeli, T., Nishino, K. (eds.) Computer Vision – ECCV 2022. vol. 13803, pp. 205--218. Springer Nature Switzerland, Cham (2023). \doi{10.1007/978-3-031-25066-8_9}

\bibitem{chen_transunet_2021}
Chen, J., Lu, Y., Yu, Q., Luo, X., Adeli, E., Wang, Y., Lu, L., Yuille, A.L., Zhou, Y.: {TransUNet}: {Transformers} {Make} {Strong} {Encoders} for {Medical} {Image} {Segmentation} (Feb 2021), \url{https://arxiv.org/abs/2102.04306v1}

\bibitem{devlin_bert_2019}
Devlin, J., Chang, M.W., Lee, K., Toutanova, K.: {BERT}: {Pre}-training of {Deep} {Bidirectional} {Transformers} for {Language} {Understanding} (May 2019). \doi{10.48550/arXiv.1810.04805}, \url{http://arxiv.org/abs/1810.04805}, arXiv:1810.04805 [cs]

\bibitem{dosovitskiy_image_2020}
Dosovitskiy, A., Beyer, L., Kolesnikov, A., Weissenborn, D., Zhai, X., Unterthiner, T., Dehghani, M., Minderer, M., Heigold, G., Gelly, S., Uszkoreit, J., Houlsby, N.: An image is worth 16x16 words: Transformers for image recognition at scale. In: International Conference on Learning Representations (2021), \url{https://openreview.net/forum?id=YicbFdNTTy}

\bibitem{gutsche_automated_2023}
Gutsche, R., Lowis, C., Ziemons, K., Kocher, M., Ceccon, G., Régio~Brambilla, C., Shah, N.J., Langen, K.J., Galldiks, N., Isensee, F., Lohmann, P.: Automated {Brain} {Tumor} {Detection} and {Segmentation} for {Treatment} {Response} {Assessment} {Using} {Amino} {Acid} {PET}. Journal of Nuclear Medicine: Official Publication, Society of Nuclear Medicine  \textbf{64}(10),  1594--1602 (Oct 2023). \doi{10.2967/jnumed.123.265725}

\bibitem{hammernik2021systematic}
Hammernik, K., Schlemper, J., Qin, C., Duan, J., Summers, R.M., Rueckert, D.: Systematic evaluation of iterative deep neural networks for fast parallel mri reconstruction with sensitivity-weighted coil combination. Magnetic Resonance in Medicine  \textbf{86}(4),  1859--1872 (2021)

\bibitem{hatamizadeh_unetr_2022}
Hatamizadeh, A., Tang, Y., Nath, V., Yang, D., Myronenko, A., Landman, B., Roth, H.R., Xu, D.: {UNETR}: {Transformers} for {3D} {Medical} {Image} {Segmentation}. 2022 IEEE/CVF Winter Conference on Applications of Computer Vision (WACV) pp. 1748--1758 (Jan 2022). \doi{10.1109/WACV51458.2022.00181}

\bibitem{huang2019brain}
Huang, Q., Chen, X., Metaxas, D., Nadar, M.S.: Brain segmentation from k-space with end-to-end recurrent attention network. In: Medical Image Computing and Computer Assisted Intervention--MICCAI 2019: 22nd International Conference, Shenzhen, China, October 13--17, 2019, Proceedings, Part III 22. pp. 275--283. Springer (2019)

\bibitem{isensee_nnu-net_2021}
Isensee, F., Jaeger, P.F., Kohl, S.A.A., Petersen, J., Maier-Hein, K.H.: {nnU}-{Net}: a self-configuring method for deep learning-based biomedical image segmentation. Nature Methods  \textbf{18}(2),  203--211 (Feb 2021). \doi{10.1038/s41592-020-01008-z}, number: 2 Publisher: Nature Publishing Group

\bibitem{jaegle_perceiver_2022}
Jaegle, A., Borgeaud, S., Alayrac, J.B., Doersch, C., Ionescu, C., Ding, D., Koppula, S., Zoran, D., Brock, A., Shelhamer, E., Hénaff, O., Botvinick, M.M., Zisserman, A., Vinyals, O., Carreira, J.: Perceiver {IO}: {A} {General} {Architecture} for {Structured} {Inputs} \& {Outputs} (Mar 2022). \doi{10.48550/arXiv.2107.14795}

\bibitem{jiang2021focal}
Jiang, L., Dai, B., Wu, W., Loy, C.C.: Focal frequency loss for image reconstruction and synthesis. In: Proceedings of the IEEE/CVF International Conference on Computer Vision. pp. 13919--13929 (2021)

\bibitem{krasser_pytorch_2023}
Krasser, M., Stumpf, C.: A {PyTorch} implementation of {Perceiver}, {Perceiver} {IO} and {Perceiver} {AR} with {PyTorch} {Lightning} scripts for distributed training. (May 2023), \url{https://github.com/krasserm/perceiver-io}

\bibitem{lee-thorp_fnet_2022}
Lee-Thorp, J., Ainslie, J., Eckstein, I., Ontanon, S.: {FNet}: {Mixing} {Tokens} with {Fourier} {Transforms} (May 2022). \doi{10.48550/arXiv.2105.03824}

\bibitem{lin_seg4reg_2022}
Lin, Y., Liu, L., Ma, K., Zheng, Y.: {Seg4Reg}+: {Consistency} {Learning} between {Spine} {Segmentation} and {Cobb} {Angle} {Regression} (Aug 2022). \doi{10.48550/arXiv.2208.12462}, arXiv:2208.12462 [cs]

\bibitem{marcus_open_2007}
Marcus, D.S., Wang, T.H., Parker, J., Csernansky, J.G., Morris, J.C., Buckner, R.L.: Open {Access} {Series} of {Imaging} {Studies} ({OASIS}): {Cross}-sectional {MRI} {Data} in {Young}, {Middle} {Aged}, {Nondemented}, and {Demented} {Older} {Adults}. Journal of Cognitive Neuroscience  \textbf{19}(9),  1498--1507 (Sep 2007). \doi{10.1162/jocn.2007.19.9.1498}

\bibitem{rao2021global}
Rao, Y., Zhao, W., Zhu, Z., Lu, J., Zhou, J.: Global filter networks for image classification. Advances in neural information processing systems  \textbf{34},  980--993 (2021)

\bibitem{singh2023emerging}
Singh, D., Monga, A., de~Moura, H.L., Zhang, X., Zibetti, M.V., Regatte, R.R.: Emerging trends in fast mri using deep-learning reconstruction on undersampled k-space data: a systematic review. Bioengineering  \textbf{10}(9), ~1012 (2023)

\bibitem{stuchi2020frequency}
Stuchi, J.A., Boccato, L., Attux, R.: Frequency learning for image classification. arXiv preprint arXiv:2006.15476  (2020)

\bibitem{tolpadi2023k2s}
Tolpadi, A.A., Bharadwaj, U., Gao, K.T., Bhattacharjee, R., Gassert, F.G., Luitjens, J., Giesler, P., Morshuis, J.N., Fischer, P., Hein, M., et~al.: K2s challenge: From undersampled k-space to automatic segmentation. Bioengineering  \textbf{10}(2), ~267 (2023)

\bibitem{touvron_resmlp_2021}
Touvron, H., Bojanowski, P., Caron, M., Cord, M., El-Nouby, A., Grave, E., Izacard, G., Joulin, A., Synnaeve, G., Verbeek, J., Jégou, H.: {ResMLP}: {Feedforward} networks for image classification with data-efficient training (Jun 2021). \doi{10.48550/arXiv.2105.03404}, arXiv:2105.03404 [cs]

\bibitem{vaswani_attention_2017}
Vaswani, A., Shazeer, N., Parmar, N., Uszkoreit, J., Jones, L., Gomez, A.N., Kaiser, L., Polosukhin, I.: Attention {Is} {All} {You} {Need} (Jun 2017), \url{https://arxiv.org/abs/1706.03762v7}

\bibitem{wang2024fremim}
Wang, W., Wang, J., Chen, C., Jiao, J., Cai, Y., Song, S., Li, J.: Fremim: Fourier transform meets masked image modeling for medical image segmentation. In: Proceedings of the IEEE/CVF Winter Conference on Applications of Computer Vision. pp. 7860--7870 (2024)

\bibitem{wu_cvt_2021}
Wu, H., Xiao, B., Codella, N., Liu, M., Dai, X., Yuan, L., Zhang, L.: {CvT}: {Introducing} {Convolutions} to {Vision} {Transformers} (Mar 2021). \doi{10.48550/arXiv.2103.15808}

\bibitem{zhang_graph_2022}
Zhang, D., Tang, J., Cheng, K.T.: Graph {Reasoning} {Transformer} for {Image} {Parsing} (Sep 2022), \url{http://arxiv.org/abs/2209.09545}, arXiv:2209.09545 [cs]

\bibitem{zhao2022k}
Zhao, Z., Zhang, T., Xie, W., Wang, Y.F., Zhang, Y.: K-space transformer for undersampled mri reconstruction. In: BMVC. p.~473 (2022)

\bibitem{zhou2022deep}
Zhou, M., Yu, H., Huang, J., Zhao, F., Gu, J., Loy, C.C., Meng, D., Li, C.: Deep fourier up-sampling. arXiv preprint arXiv:2210.05171  (2022)

\end{thebibliography}

\appendix
\newpage
\section{Appendix}
\begin{figure}[h!]
\centering
\includegraphics[width=0.65\textwidth]{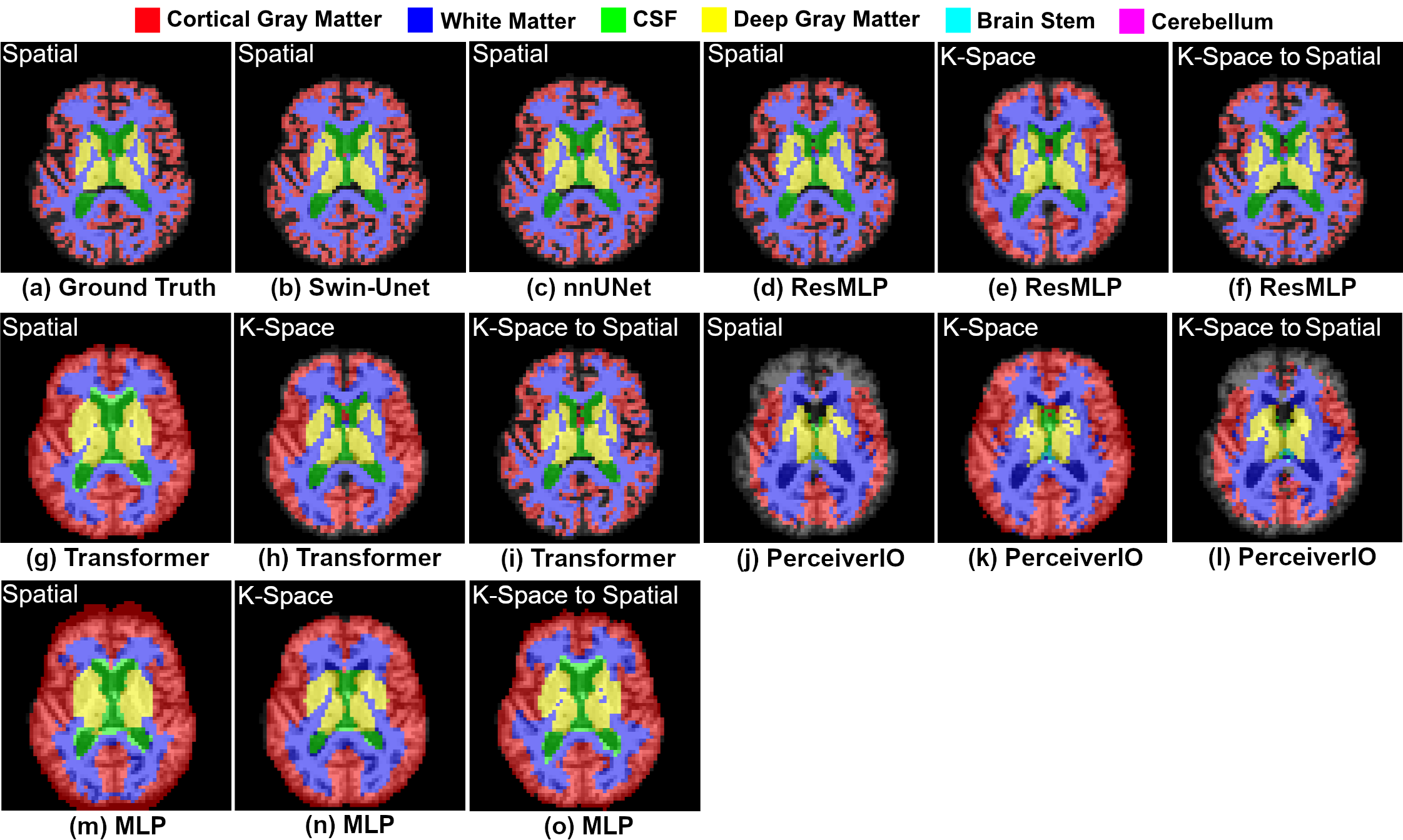}
\caption{Segmentation masks of the different models for brain tissue segmentation among varies domains.}
\label{fig:all_tissue}
\end{figure}

\begin{table}
    \small
    \centering
    \begin{tabularx}{\textwidth}{lXXXXX}
        \toprule
        Domain & \multicolumn{5}{c}{Spatial} \\
        \cmidrule(lr){2-6}
        Anatomy & Metric & MLP & ResMLP & PerceiverIO & Transformer \\
        \midrule
        CSF & Dice & 0.562 & \textbf{0.868} & 0.078 & 0.686 \\
        \midrule
        Cortical Gray Matter & Dice & 0.491 & \textbf{0.773} & 0.263 & 0.558 \\
        \midrule
        White Matter & Dice & 0.602 & \textbf{0.874} & 0.470 & 0.673 \\
        \midrule
        Deep Gray Matter & Dice & 0.645 & \textbf{0.828} & 0.317 & 0.699 \\
        \midrule
        Brain Stem & Dice & 0.506 & \textbf{0.883} & 0.405 & 0.589 \\
        \midrule
        Cerebellum & Dice & 0.623 & \textbf{0.916} & 0.443 & 0.723 \\
        \midrule
        \multirow{3}{*}{All} & Dice & 0.625 & \textbf{0.876} & 0.418 & 0.698 \\
        & Recall & 0.829 & 0.879 & 0.379 & 0.864 \\
        & Specificity & 0.979 & 0.987 & 0.926 & 0.984 \\
        \bottomrule
    \end{tabularx}
    \caption{Performance metrics in the spatial domain for different models on brain tissue segmentation.}
    \label{tab:ex1_tissue_spatial}
\end{table}
\begin{table}
    \small
    \centering
    \begin{tabularx}{\textwidth}{lXXXXX}
        \toprule
        Domain & \multicolumn{5}{c}{K-Space} \\
        \cmidrule(lr){2-6}
        Anatomy & Metric & MLP & ResMLP & PerceiverIO & Transformer \\
        \midrule
        CSF & Dice & 0.570 & \textbf{0.844} & 0.098 & 0.795 \\
        \midrule
        Cortical Gray Matter & Dice & 0.522 & \textbf{0.581} & 0.395 & 0.561 \\
        \midrule
        White Matter & Dice & 0.612 & \textbf{0.723} & 0.486 & 0.688 \\
        \midrule
        Deep Gray Matter & Dice & 0.668 & \textbf{0.806} & 0.333 & 0.774 \\
        \midrule
        Brain Stem & Dice & 0.752 & \textbf{0.880} & 0.471 & 0.860 \\
        \midrule
        Cerebellum & Dice & 0.745 & \textbf{0.892} & 0.510 & 0.874 \\
        \midrule
        \multirow{3}{*}{All} & Dice & 0.690 & \textbf{0.815} & 0.461 & 0.790 \\
        & Recall & 0.757 & 0.811 & 0.507 & 0.788 \\
        & Specificity & 0.979 & 0.975 & 0.960 & 0.974 \\
        \bottomrule
    \end{tabularx}
    \caption{Performance metrics in k-space for different models on brain tissue segmentation.}
    \label{tab:ex1_tissue_kspace}
\end{table}
\begin{table}
    \small
    \centering
    \begin{tabularx}{\textwidth}{lXXXXX}
        \toprule
        Domain & \multicolumn{5}{c}{K-Space $\rightarrow$ Spatial} \\
        \cmidrule(lr){2-6}
        Anatomy & Metric & MLP & ResMLP & PerceiverIO & Transformer \\
        \midrule
        CSF & Dice & 0.645 & \textbf{0.866} & 0.054 & 0.853 \\
        \midrule
        Cortical Gray Matter & Dice & 0.539 & \textbf{0.801} & 0.270 & 0.761 \\
        \midrule
        White Matter & Dice & 0.645 & \textbf{0.898} & 0.464 & 0.858 \\
        \midrule
        Deep Gray Matter & Dice & 0.686 & \textbf{0.826} & 0.323 & 0.804 \\
        \midrule
        Brain Stem & Dice & 0.568 & \textbf{0.880} & 0.410 & 0.863 \\
        \midrule
        Cerebellum & Dice & 0.696 & \textbf{0.919} & 0.466 & 0.902 \\
        \midrule
        \multirow{3}{*}{All} & Dice & 0.676 & \textbf{0.883} & 0.419 & 0.861 \\
        & Recall & 0.856 & 0.881 & 0.385 & 0.865 \\
        & Specificity & 0.983 & 0.987 & 0.927 & 0.986 \\
        \bottomrule
    \end{tabularx}
    \caption{Performance metrics the in k-space-to-spatial domain for different models on brain tissue segmentation.}
    \label{tab:ex1_tissue_kspatial}
\end{table}

\end{document}